\documentstyle{trigger12}

\input epsf

\begin{document}

\begin{frontmatter}

\title{Characteristics of the Multi-Telescope Coincidence Trigger
of the HEGRA IACT System}

\author{N. Bulian},
\author{A. Daum},
\author{G. Hermann},
\author{M. He{\ss}},
\author{W. Hofmann},
\author{H. Lampeitl},
\author{G. P\"uhlhofer},
\author{C. K\"ohler},
\author{M. Panter},
\author{M. Stein}

\address{Max-Planck-Institut f\"ur Kernphysik, P.O. Box 103980,
        D-69029 Heidelberg, Germany}

\author{G. B\"orst},
\author{G. Rauterberg},
\author{M. Samorski},
\author{C. Sauerland},
\author{W. Stamm}

\address{Universit\"at Kiel, Inst. f\"ur Kernphysik,
       Olshausenstr.40, D-24118 Kiel, Germany}

\begin{abstract}
The HEGRA--collaboration is operating a system of imaging atmospheric
Cherenkov telescopes to search for sources of TeV--$\gamma$--rays.
 Air showers are observed in
stereoscopic mode with several telescopes simultaneously. To trigger
the telescope system a versatile two--level trigger scheme has been
implemented, which allows a significant reduction of the energy threshold
with respect to single telescopes. The technical implementation of this
trigger scheme and the performance of the trigger system
are described. Results include the dependence of single- and multi-telescope
trigger rates on the trigger thresholds, on the orientation of the 
telescopes, and on the type of the primary particle.
\end{abstract}

\end{frontmatter}

\section{Introduction}

Imaging atmospheric Cherenkov telescopes (IACT) \cite{weekes} have proven to be
the most powerful instruments in TeV--$\gamma$--ray astronomy.
The high sensitivity of these instruments results from the large effective
collection area, governed by the size of the light pool of air showers,
and from an effective hadron suppression
\cite{fegan}, achieved by the imaging technique.
Further improvements of the Cherenkov technique can be obtained
by coupling several telescopes to a system for stereoscopic observations
of air showers. 

The HEGRA-collaboration has installed such a stereoscopic
system during the last two years and first measurements show very encouraging
results~\cite{system_crab,system_501}. 
The HEGRA IACT system is located on the Canary Island of La Palma, 
at the Observatorio del Roque de los Muchachos
of the Instituto Astrofisico de Canarias,
at a height of about 2200~m asl.
In its final form, the HEGRA IACT array will consist of 
five identical telescopes with 8.5~m$^2$ mirror area, 
5~m focal length, and 271-pixel cameras~\cite{hermann_padua} with a pixel
size of $0.25^\circ$ and a field of view of $4.3^\circ$.
Four telescopes (CT2, CT4, CT5, CT6)
 are arranged in the corners of a square with 
roughly
100~m side length, and one telescope (CT3) is located 
in the center of the square. At this time, only four of the
telescopes (CT3, CT4, CT5, CT6)
are included in the IACT system; the last telescope
(CT2, at one corner of the square) is still equipped with an older
camera and is operated independently.
The cameras are read out by Flash-ADCs with 120 MHz sampling rate and
34~$\mu$s memory depth.

Triggering of such an IACT system is rather more complex than for single IACTs.
Given that the stereoscopic analysis requires good images in at
least two telescopes, already at the trigger level
a multi-telescope coincidence is requested. Such a coincidence trigger
will very effectively suppress background triggers generated by
the light of the night sky and by local muons, allowing to significantly
reduce the trigger threshold and hence the energy threshold. A telescope
coincidence is also expected to enhance the ratio of $\gamma$--ray
triggers to cosmic-ray events.

In this paper, we give an overview of the
characteristics of the trigger of the HEGRA IACT array, with
emphasis on trigger rates and their dependence on the various
parameters; the technical implementation is also summarized briefly.

\section{The two--level trigger system}

The trigger system of the HEGRA IACTs is designed as a two-level
system. First, for each telescope a `local' trigger is derived by requiring
that signals in a minimum number of pixels (typically 2) exceed
a given threshold, optionally imposing additional topological 
constraints emphasizing the compactness of shower images. 
The pattern of trigger pixels is latched for later use. 
This first local trigger level
operates essentially without deadtime, allowing in extreme cases local
trigger rates in the MHz range. The local trigger electronics is
housed in a container near each telescope, together with the
Flash-ADC digitizers and a CPU for readout and data buffering
as well as for telescope control and monitoring. 
The local telescope trigger signals are routed to a central 
station. There,
in the second `global' trigger level, several telescopes
are required to trigger in coincidence. During the about
1 $\mu$s required for the propagation and processing of trigger signals,
event data are stored locally
in the Flash-ADC memory; only after a global
trigger, the continuous conversion is stopped and data are read out.
Since the full time history of the last 34~$\mu$s for all pixels
of all telescopes is stored, event data can be collected irrespectively
of whether a given telescope contributed to the global trigger or not.

To provide maximum flexibility, all relevant parameters, 
such as the pattern of pixels included in
the trigger, the thresholds, and the coincidence levels are under computer control.
This feature allows the trigger system to automatically adapt
to changing conditions, such as bright stars moving through the
field of view. The extensive use of configuration files and
history files ensures that trigger conditions are well documented
and reproducible.

The local trigger electronics at each telescope 
relies primarily on three different
types of logic modules: the
Discriminator--Monitor Cards (DMC),
the majority unit, and the topological trigger unit.

\begin{figure}[htb]
\begin{center}
\mbox{
\epsfxsize9.0cm
\epsffile{ 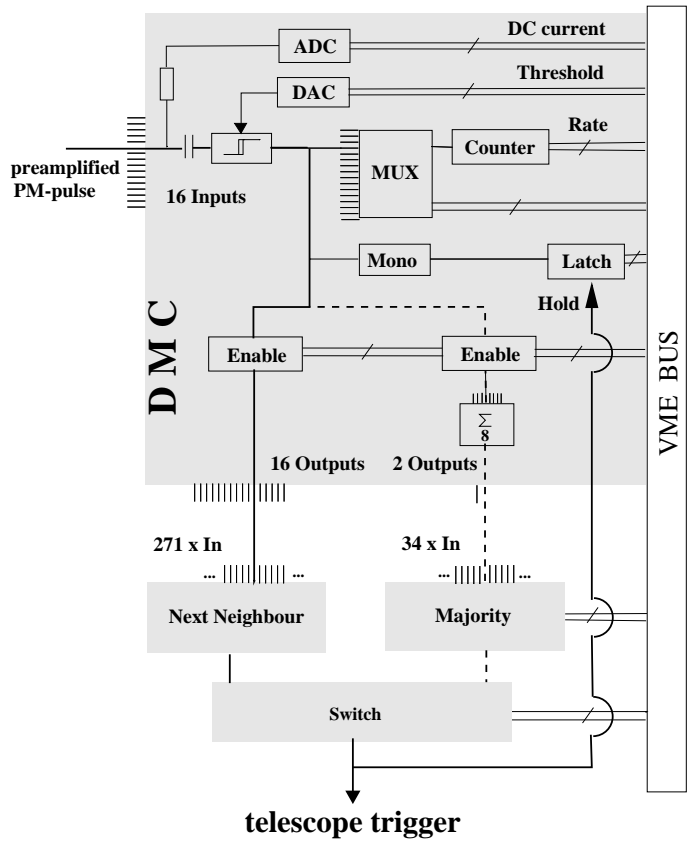}
}
\end{center}
\caption
{Configuration of the trigger electronics, emphasizing the function
of the Discriminator-Monitor Cards (DMCs), 
with the trigger discriminators, gated trigger
outputs and latches, and monitoring of PMT currents as well as
trigger rates.}
\label{fig_dmc}
\end{figure}

Signals from the camera PMTs are amplified in the camera by a 
factor 16 and are then routed via 22~m of RG178-cable to the DMCs
and the Flash-ADCs. The amplifiers in the camera also serve to 
convert the average DC current from the PMTs into DC voltage levels,
which are superimposed to the AC-coupled signals~\cite{hermann_padua}. 
Typical PMT signals arriving
at the DMC cards have a signal level of 1.2~mV/photoelectron and 6~ns pulse
width (FWHM).
Each DMC (Fig.~\ref{fig_dmc}) serves 16 analog input channels. For each 
channel, it provides an input amplifier of gain 3 and a  discriminator.
Discriminator
thresholds can be set independently via VME, with a threshold up to
80 mV (referring to the level at the DMC input), 
equivalent to a range of roughly 0 to 70 photoelectrons. 
Discriminator outputs are available both individually on front-panel 
connectors, and as 8-channel summed outputs,
which provide a voltage proportional to the number of triggered pixels. 
Channels can be individually 
disabled via VME. This feature is used to exclude
single channels dynamically from the trigger 
if, e.g., a bright star is imaged into the pixel. The length of the
discriminator output signals is hardware-adjustable, and is set to 17~ns. 
In order to preserve the information on triggered pixels, the discriminator
outputs can be latched on the DMC, by applying an external `hold' 
signal when the telescope triggers. Since the latch signal will only 
arrive with a certain delay caused by the trigger electronics and cables,
the input signals to the VME-readable
latches are extended to 100 ns length by a monoflop, to guarantee
sufficient overlap. The DMC contains
furthermore a
multiplexed scaler to monitor pixel rates, and ADCs for the measurements of the 
average PMT currents.

The 8-channel summed outputs from the DMCs - 34 in total to serve 
the 271 pixels of a camera - are routed to the 
majority unit, where the outputs of all DMCs are summed up and discriminated with
a software adjustable threshold, corresponding to the minimal number of
pixels required for the trigger. A minimal overlap of the input signals of
3~ns is needed, resulting  in an effective coincidence window of 
14~ns. The usual trigger condition requires the coincidence of two pixels;
Monte Carlo studies showed that for the given pixel
size of $0.25^\circ$, 
such a two-pixel coincidence is more efficient in rejecting backgrounds
than, say, a three-pixel
coincidence with lower pixel thresholds.
The majority unit again contains a scaler
to monitor trigger rates, and additional inputs for auxiliary triggers,
which can be enabled under VME control.

An alternative `next-neighbor' (NN) trigger is provided 
by the topological trigger unit, which receives as input the 271 
individual discriminator outputs, and 
which requires at least two triggering pixels to be neighbors. The logic
is implemented using programmable gate arrays; the typical trigger
decision time of 20~ns is comparable to that of the majority unit. 
The required minimum
overlap of the input signals is 3~ns.
With this logic,
accidental background triggers, caused by the light of the night sky,
are reduced by a factor of about 48, while over 95\% of
the compact $\gamma$-ray images are kept~\cite{nn_paper}. The 
improved suppression of accidental triggers allows to reduce pixel
trigger thresholds and hence the energy threshold of the telescope. 

The NN trigger signal is fed back into one of the additional trigger 
inputs of the majority unit; via VME, either this external trigger
or the internal majority signal can be switched to the trigger outputs,
which also generate the hold signals for the latch registers on the DMCs. 
For normal observations, the NN trigger
is enabled; the majority trigger is nowadays
used only for special studies and
tests.

\begin{figure}[htb]
\begin{center}
\mbox{
\epsfxsize9.0cm
\epsffile{ 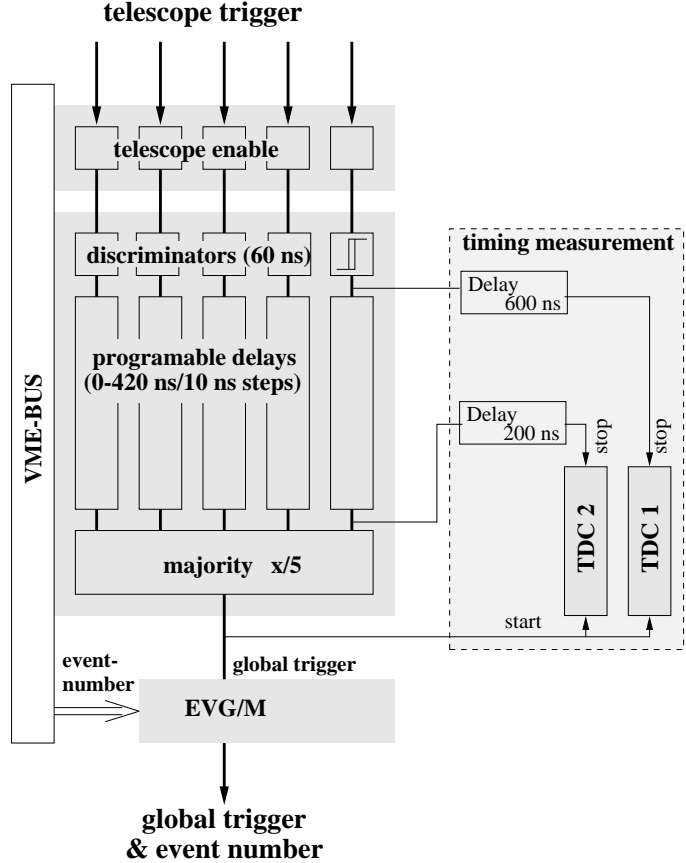}
}
\end{center}
\caption
{Sketch of the central trigger logic of the HEGRA IACT array,
and of the TDCs used to monitor trigger timing.
The modules and connections used for the calibration 
of the TDCs are not shown.}
\label{fig_central}
\end{figure}
The local trigger signals of each telescope are routed to the 
central station using cables of identical length. There,
the global trigger for the IACT system is derived.

\begin{figure}[htb]
\begin{center}
\mbox{
\epsfxsize9.0cm
\epsffile{ 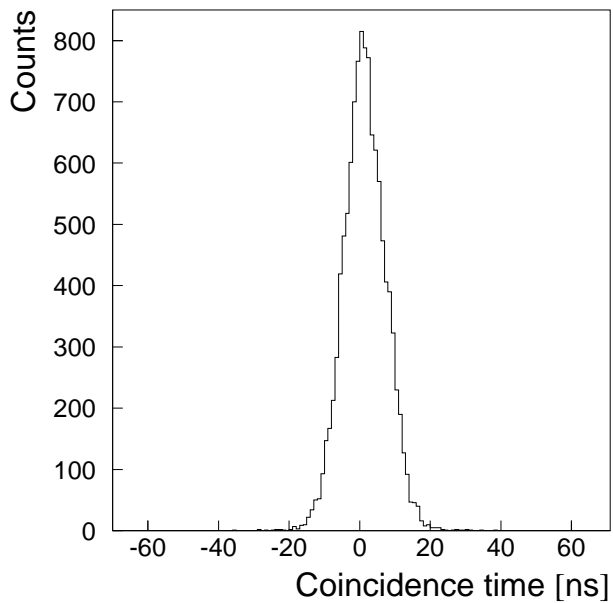}
}
\end{center}
\caption
{Distribution in the differences of the arrival times of triggers
from the two telescopes CT3 and CT5, after approximate compensation of the 
differences in propagation delay. The spacing between the telescopes is
75~m.}
\label{fig_tdc}
\end{figure}

The main task of the central station is to generate a trigger
if at least a preset number of telescopes, typically two, have triggered
within a given coincidence time. In order to allow a short window 
for these inter-telescope coincidences, differences in the arrival times
of the Cherenkov light front at the individual telescopes have to be
compensated. With a separation of up to 140~m between telescopes, 
arrival times of trigger signals 
may differ by almost 400~ns for showers at angles of 
$60^\circ$ from  the zenith. Key element of the central station is 
therefore a VME board with five computer-controlled delay units, feeding
a majority discriminator (Fig. \ref{fig_central}). The delays are 
implemented as switched delay lines, 
adjustable in steps of 10 nsec. The delay settings are updated
continuously while a source is tracked across the sky. The width of the
inter-telescope coincidence window is set to a conservative value of 70~ns.
Trigger arrival times both before and after the delay are monitored 
with TDCs. Fig.~\ref{fig_tdc} shows the relative timing of trigger 
signals from two telescopes (CT3 and CT5), 
with a width of about 6~ns rms. Apart from
the finite step size of the delay adjustments, a 
number of effects contribute to this width. The intrinsic time resolution
of the telescope signals is of order 1~ns and is negligible in this 
context. More important is the fact that the axes of triggered showers 
may deviate by as much as $2^\circ$ from the telescope axis (limited 
essentially by field of view of the cameras). Correspondingly, 
given the 75~m spacing of the two telescopes, arrival
delays may differ by up to 7~ns from their nominal values. In 
addition, since the Cherenkov light front resembles a spherical
wave front rather than a plane wave, 
the relative timing may vary by another
5~ns depending on the location of the triggered telescopes relative to
the shower core.

The global trigger signal is then sent to all telescopes of the system
 by a custom unit (EVG/M in Fig.~\ref{fig_central}), 
which appends to the basic trigger signal an 
event number used to synchronize events. This unit also provides the option
to calibrate the cable delays between the central station and the telescopes.
The receiver units trigger, via an interrupt, the readout of the Flash-ADCs
by the local CPUs at each telescope. During interrupt handling, further 
triggers are disabled, but the receiver units keep count of global
triggers arriving 
during this deadtime. Since the processing time of individual telescopes
may vary, the deadtime of the telescopes it largely, but not completely 
correlated. At typical trigger rates of 15~Hz, the global system dead time 
is below 5 \%.  About 97\% of the events are complete in the sense that
all telescopes are read out; in the remaining 3\%, at least one telescope
was still busy processing previous triggers.

\section{Cosmic-ray trigger rates}

As part of the commissioning of the IACT system, trigger rates at the 
various levels - pixel rates, local telescope trigger rates, and 
global system trigger
rates - were studied as a function of pixel trigger thresholds. For these 
measurements, the telescopes were pointed at a dark region of the sky
near the zenith. Using a pulsed light source
in the center of the mirror dish, 
the cameras were flat-fielded by adjusting the high-voltage of the individual
PMTs. To determine offset voltages of 
the discriminators, thresholds were scanned, watching the discriminators
toggle once the threshold crossed the offset. Offsets determined by this
technique vary by a few mV; discriminator thresholds quoted
in the following are offset-corrected.

From the width of the distribution of digitized amplitudes of the
laser calibration pulses one can, after corrections for the width of
the single-photoelectron peak and for intensity fluctuations of the
laser, determine the average number of photoelectrons per pixel, and hence
the pulse height equivalent to a single photoelectron.
An average single-photoelectron signal corresponds to a 
voltage at the discriminator of about 1.2~mV. We note, however, that the
PMTs are adjusted to provide a fixed signal for a given (relatively large)
amount of light; this means that the product of PMT quantum efficiency, 
collection efficiency, PMT gain and electronics gain is constant, but 
not necessarily the PMT gain and electronics gain alone, which determine the
single-photoelectron signal.

\begin{figure}[htb]
\begin{center}
\mbox{
\epsfxsize9.0cm
\epsffile{ 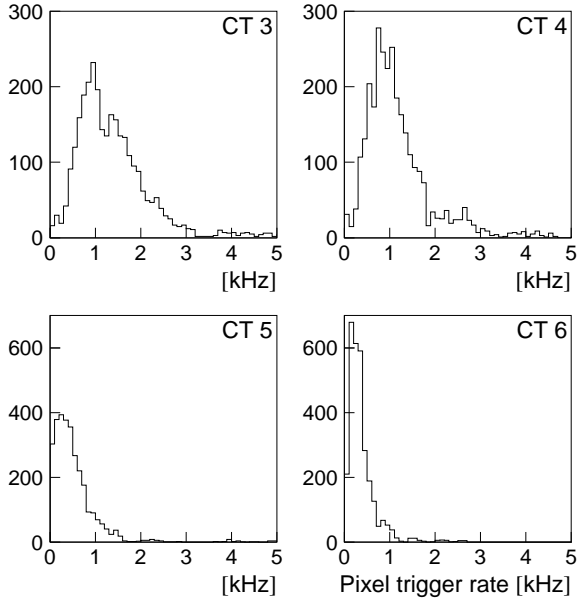}
}
\end{center}
\caption
{Distribution of the trigger rate of individual pixels at a 
fixed pixel threshold of 8~mV, for each of the 
telescopes. For each pixel, multiple rate measurements 
are contained in the plot.}
\label{fig_pix_rates}
\end{figure}

\begin{figure}[htb]
\begin{center}
\mbox{
\epsfxsize9.0cm
\epsffile{ 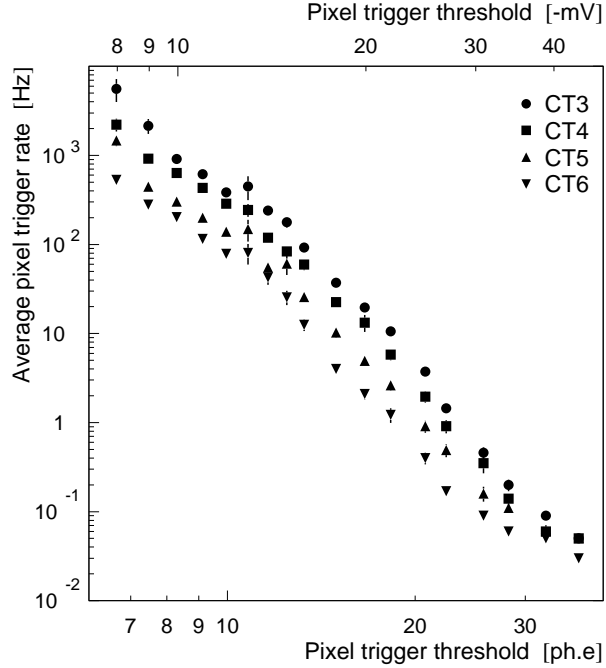}
}
\end{center}
\caption
{Trigger rate of single pixels, averaged over all pixels of a telescope,
as a function of the trigger threshold both in mV and in photoelectrons.}
\label{fig_pixel}
\end{figure}

Fig.~\ref{fig_pix_rates} shows the distribution of pixel trigger rates for
each of the CT cameras, for a pixel trigger threshold of 8~mV.
Rates vary significantly 
both from pixel to pixel, and between telescopes. 
As discussed elsewhere
\cite{trigger_paper,munich_paper}, the prime source of single-pixel triggers
under typical IACT conditions are not genuine multi-photoelectron events
(the rate of which should be pixel-independent), but instead large 
afterpulses of single-photoelectron events, generated when an atom of the
residual gas in the PMT is ionized and accelerated towards the photocathode,
where it knocks out further electrons. 
The rate of such afterpulses varies
strongly between individual PMTs.
Telescopes CT3 and CT4 differ from CT5 and CT6 
by about a factor of 3 in the mean pixel rate
at a given threshold; this difference is not too surprising, since
PMTs were grouped according to the high voltage required for a given
gain, and the cameras of CT3 and CT4 require slightly higher average
operating voltages, compared to CT5 and CT6. 
The afterpulse rate, on the other hand, depends
strongly on the voltage applied between photocathode 
and first dynode \cite{trigger_paper}.
Fig.~\ref{fig_pixel} shows the average pixel trigger rates of the telescopes
CT3 through CT6, as a function of the trigger thresholds. 

\begin{figure}[tb]
\begin{center}
\mbox{
\epsfxsize8.0cm
\epsffile{ 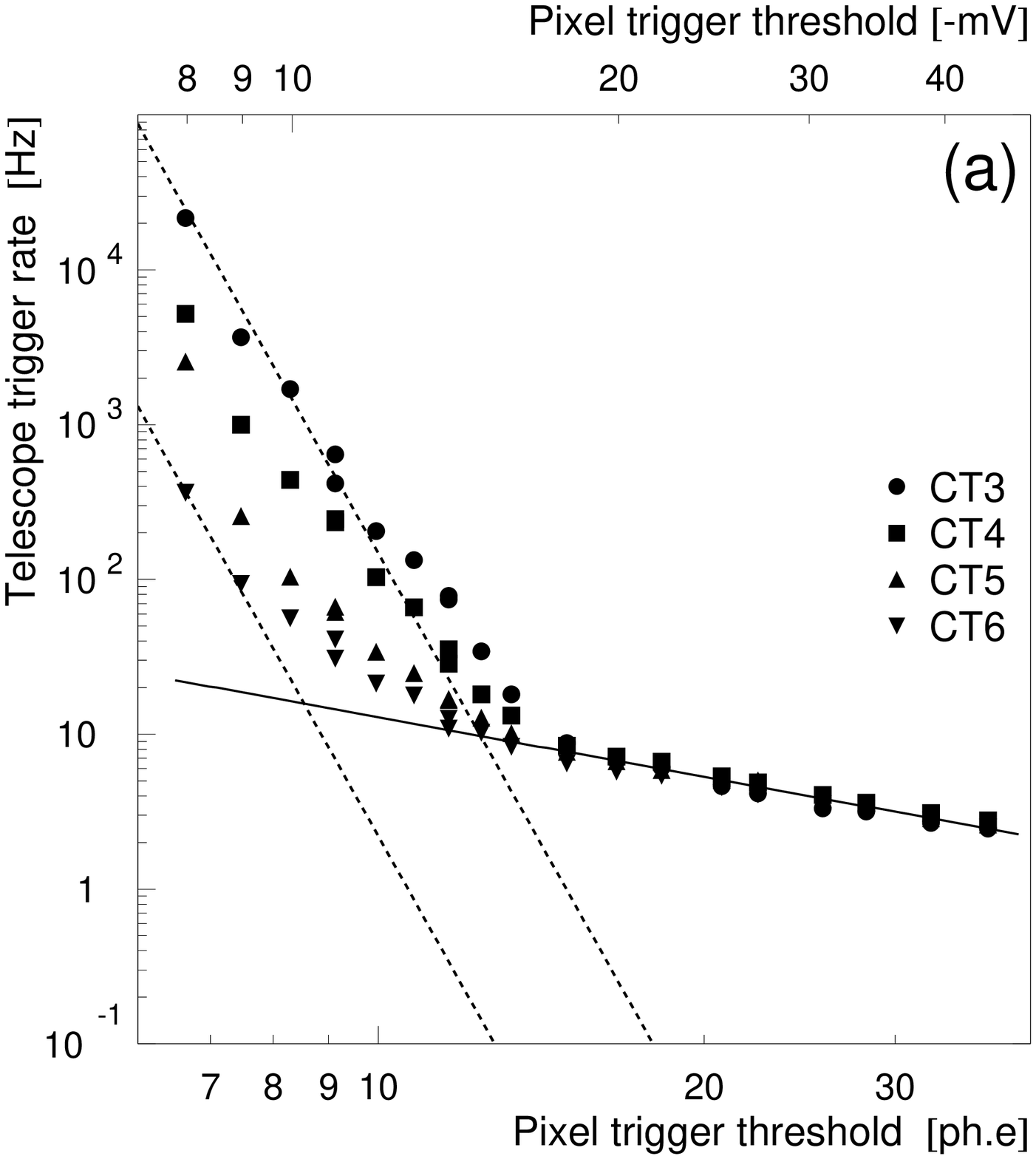}
}
\mbox{
\epsfxsize8.0cm
\epsffile{ 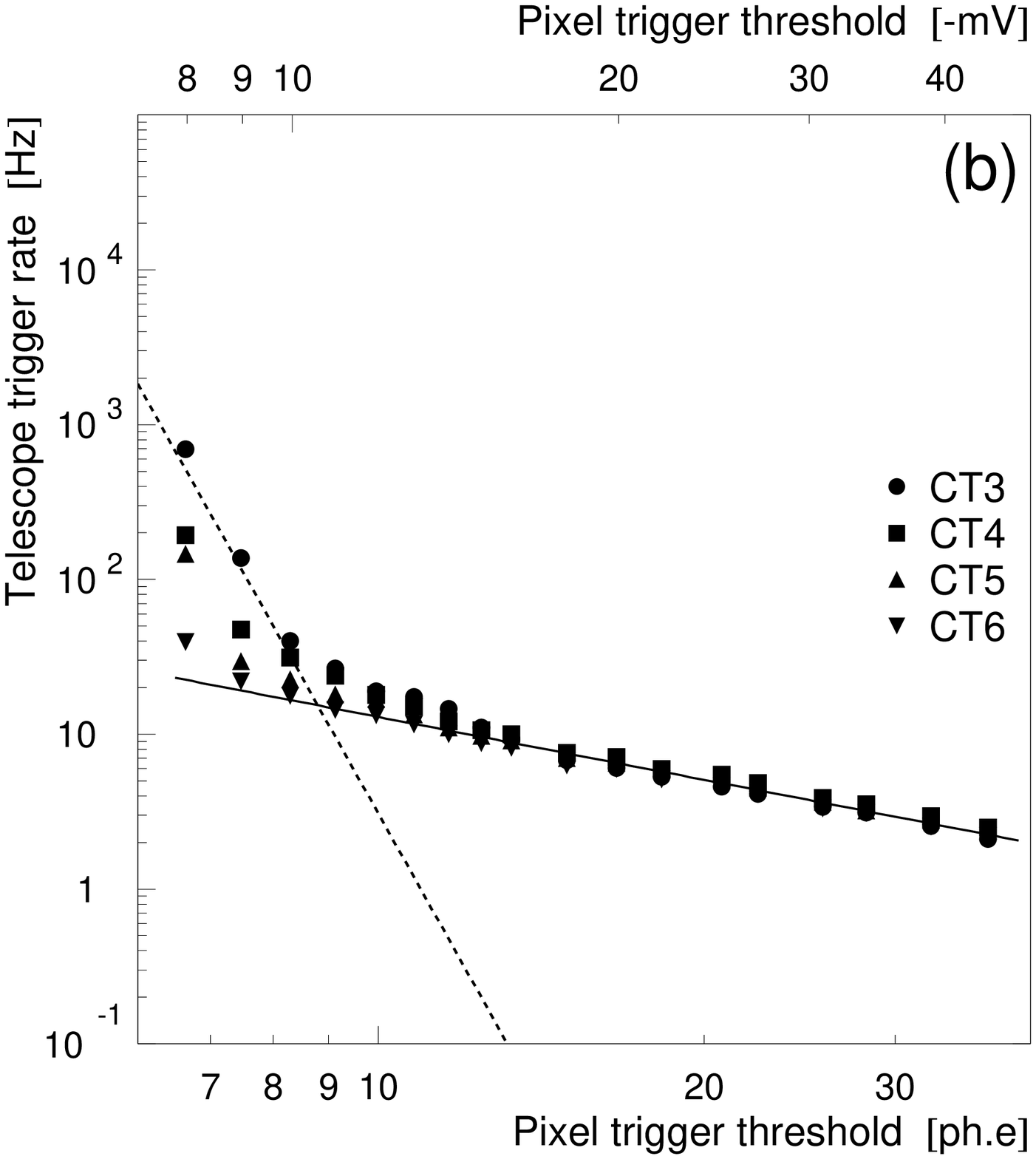}
}
\end{center}
\caption
{(a) Two-pixel coincidence rate of the different telescopes,
as a function of the trigger threshold both in mV and in photoelectrons.
Also shown are the random-coincidence rates for CT3 (upper 
dashed line) and CT6 (lower dashed line)
calculated from the measured pixel trigger rates assuming a coincidence 
time of 14 ns. The full line represents a power-law fit to the data
points for larger thresholds.
(b) Same for the next-neighbor
trigger, which requires that at least two of the triggered pixels
are direct neighbors. The dashed line indicates the random rate for CT3.}
\label{fig_tel}
\end{figure}

The two-pixel coincidence rate is displayed in Fig.~\ref{fig_tel}(a),
for the majority trigger unit.
It displays two separated regimes. At high thresholds, above about 18~mV,
or 15 photoelectrons, random pixel coincidences are rare and the 
rate of events is determined by genuine cosmic-ray triggers. The trigger
rates follow a power law spectrum (see Table~\ref{tab_rate}), 
and all telescopes behave almost identical.
Small rate differences between telescopes 
at the level of 10\% are most likely related to
the differences in the age and hence reflectivity of the mirrors (measured
reflectivities vary between 81\% and 85\% \cite{rasmik_refl}), and also
to differences in the quality of the mirror alignment. 
The power-law slope of the rate curve is slightly smaller than the
cosmic-ray integral spectral index; this behaviour is reproduced by
Monte Carlo simulations and results from the slightly nonlinear
relation between the shower energy and the density of Cherenkov light on
the ground, and from the influence of fluctuations. 
At lower thresholds,
coincidence rates rise steeply with decreasing thresholds, and are dominated
by random coincidences. In this regime, significant differences between 
the telescopes reflect the differences in the mean PMT trigger rates;
the measured rates are consistent with the expected random rates, as
derived using the measured pixel trigger rates and a 14~ns coincidence 
window.
\begin{table} [htb]
\begin{center}
\begin{tabular}{|l|c|c|}
\hline
Trigger condition & Rate at 15~mV pixel threshold & Spectral index \\
\hline 
Single telescope, majority & $9.7 \pm 1.3$ Hz & $1.28 \pm 0.18$ \\
Single telescope, NN trigger & $9.6 \pm 0.8$ Hz & $1.35 \pm 0.09$ \\
\hline
$\ge 2$ telescopes, majority & 7.9 Hz & 1.45 \\
$\ge 3$ telescopes, majority & 4.1 Hz & 1.35 \\
$    4$ telescopes, majority & 1.7 Hz & 1.42 \\
\hline
$\ge 2$ telescopes, NN trigger & 7.2 Hz & 1.33 \\
$\ge 3$ telescopes, NN trigger & 3.8 Hz & 1.34 \\
$    4$ telescopes, NN trigger & 1.6 Hz & 1.45 \\
\hline
\end{tabular}
\vspace{0.5cm}
\caption{Trigger rate and effective spectral index for different
trigger conditions, derived using a power-law fit to the measured
rates as a function of threshold in the regions where rates are
dominated by genuine air showers. Rates quoted represent the fit value
for a common pixel threshold of 15~mV. The errors quoted for the 
single-telescope values represent the variation of values between
telescopes. Statistical errors on the rates are generally below 0.1 Hz, and 
errors on the power-law slope are below 0.05.}
\label{tab_rate}
\end{center}
\end{table}

The next-neighbor (NN) trigger unit should effectively suppress random
coincidences by restricting the number of possible pixel combinations.
Fig. ~\ref{fig_tel}(b) shows the corresponding rates. Indeed, the 
onset of a significant rate of random triggers occurs at significantly
lower thresholds of around 10 mV to 15 mV, or 8 to 12 photoelectrons. 
For high thresholds,
trigger rates are close to the rates with the 
majority unit (see also Table~\ref{tab_rate}),
indicating minimal losses ($\le$ 10\%) even for the more diffuse proton images.

\begin{figure}[tb]
\begin{center}
\mbox{
\epsfxsize8.0cm
\epsffile{ 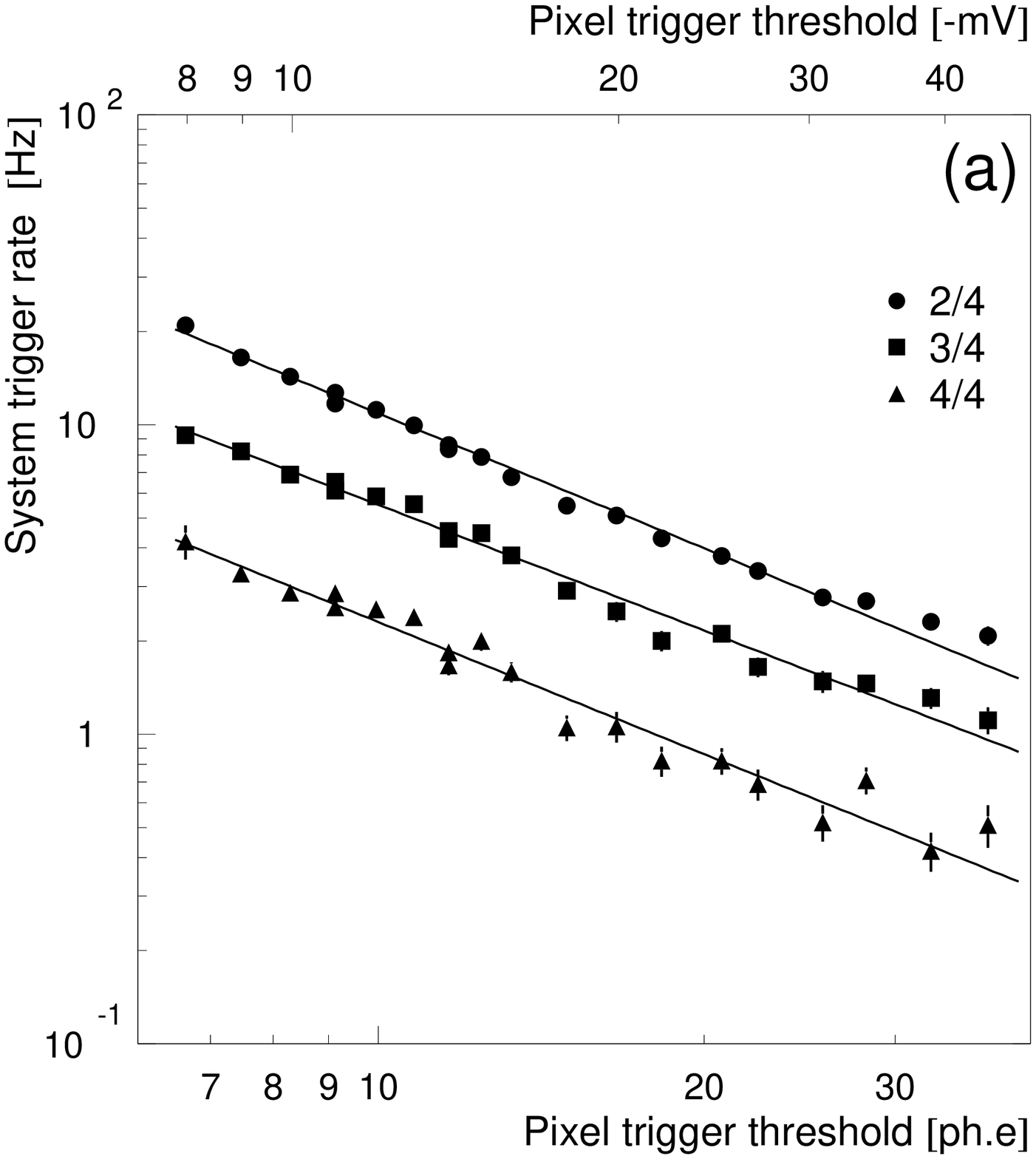}
}
\mbox{
\epsfxsize8.0cm
\epsffile{ 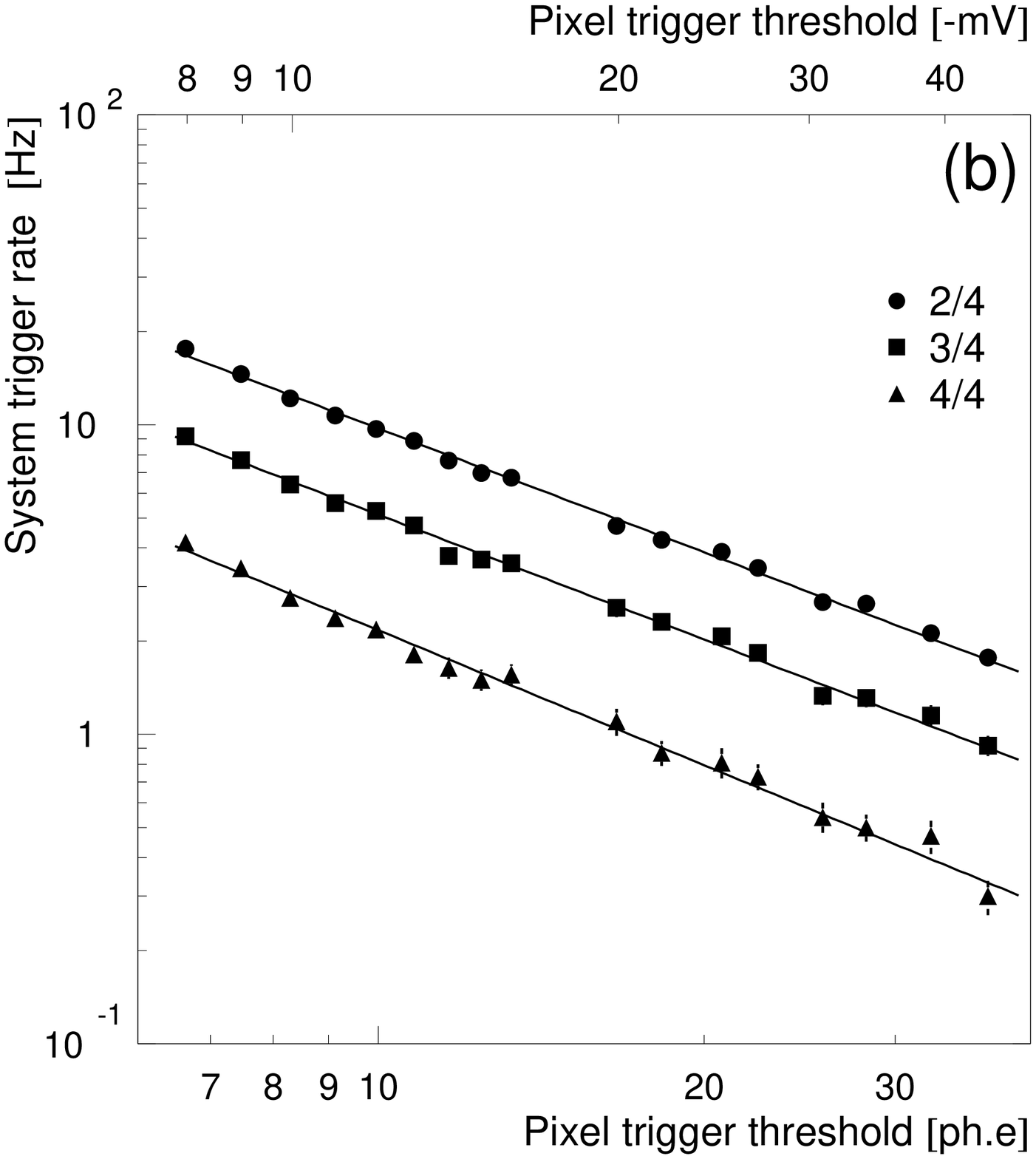}
}
\end{center}
\caption
{(a) Two-, three- and four-telescope coincidence rate,
as a function of the trigger threshold both in mV and in photoelectrons,
using the any-two-pixel (majority) trigger to derive telescope triggers.
(b) Same using the next-neighbor (NN)
trigger.}
\label{fig_array}
\end{figure}

The any-two-pixel or next-neighbor coincidence signals are sent to the central
trigger station, which checks for coincidences with other telescopes.
Two-, three- and four-telescope trigger rates are summarized in
Fig.~\ref{fig_array}, 
for the two-pixel majority trigger and for the 
next-neighbor trigger. For all conditions studied, there is no
evidence for the onset of a significant contribution of random
triggers. Rate estimates show, indeed, that for the two-telescope,
any-two-pixel trigger the rate of random telescope coincidences
amounts to about 30\% of the total rate at the lowest threshold value;
for the next-neighbor logic the random fraction drops to about 
a fraction of a percent.
With such a multi-telescope coincidence, random coincidences
do no longer play a significant role in the choice of trigger
thresholds, and other factors enter, such as the stability of
the electronics for such low thresholds, and the 
fact that images with only a few 10 photoelectrons provide 
limited information. Presently, the HEGRA IACT system is operated
with 8~mV pixel thresholds and a two-telescope coincidence 
requirement.

The rates and effective spectral index
for the different cases are included in Table~\ref{tab_rate}.
The spectral index, the ratios between 1, 2, 3 and 4-telescope
coincidence rates and -- within the 10\% to 20\% uncertainty in the
sensitivity of the telescopes -- also the absolute rates are
reproduced by Monte Carlo simulations; a detailed comparison
of measured and predicted rates and of event characteristics will
be the subject of a forthcoming paper.

%
%

\section{Dependence of rates on telescope alignment}

In a stereoscopic system of IACTs, the relative orientation of
the telescopes enters as an additional parameter. Two options are
frequently discussed:
\begin{itemize}
\item one can either align the axes of all telescopes parallel, or
\item recognizing that most of the Cherenkov light emerges 
around the shower maximum, at a finite height of
$h \approx 6$~km above the telescopes
(for typical HEGRA conditions), one can tilt the telescopes 
inward such that they point to the same spot at the height of
the shower maximum.
\end{itemize}
In a somewhat simplified picture, 
two conditions need to be fulfilled for a telescope to trigger:
the telescope needs to be located in the Cherenkov light pool, and
the shower maximum needs to be within the field of view. 
Tilting two telescopes towards each other
by about $d/h \approx 0.7^\circ$ to $1^\circ$ should provide the
maximum overlap of the fields of view and hence the 
maximum trigger rate. Here, $d \approx 70$~m to 100~m is the telescope 
spacing.

\begin{figure}[htb]
\begin{center}
\mbox{
\epsfxsize9.0cm
\epsffile{ 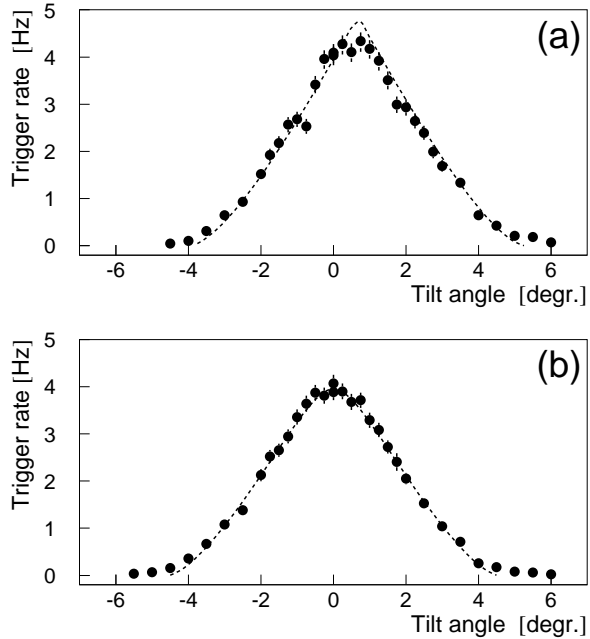}
}
\end{center}
\caption
{Coincidence rates between telescopes CT3 and CT4 as a function
of the relative alignment of the telescope axes. In (a), the 
axes are tilt along the direction connecting the telescopes; a
positive tilt means that the telescopes look towards each other.
In (b), the telescopes are tilt perpendicular to the connecting
line. The dashed curves correspond to the geometrical overlap
of the camera fields of view at a height of 6.3~km above the observation
level.}
\label{fig_tilt}
\end{figure}

To verify this expectation, coincidence rates of the two telescopes 
CT3 and CT4 were measured as a function of their relative pointing.
Fig.~\ref{fig_tilt}(a) shows the coincidence rate as a function
of the tilt along the direction connecting the two telescopes.
One finds indeed that the maximum rate is achieved when the
telescopes are slightly canted towards each other; the dependence
of coincidence rates as a function of the tilt is reproduced 
remarkably well by a simple geometrical model, calculating the
overlap of the circular fields of view at the height of the shower maximum
(dashed line). A tilt perpendicular to the connecting line
of course yields a symmetrical curve (Fig.~\ref{fig_tilt}(b)).

One also notices that the relative alignment of the telescopes
is not particularly critical; from the point of view of trigger
rates, deviations up to a fraction of a degree are tolerable.
Given the modest resulting losses, the HEGRA system telescopes
have so far been operated with parallel axes, for operational
convenience.

\section{Dependence of rates on the zenith angle}

\begin{figure}[htb]
\begin{center}
\mbox{
\epsfxsize9.0cm
\epsffile{ 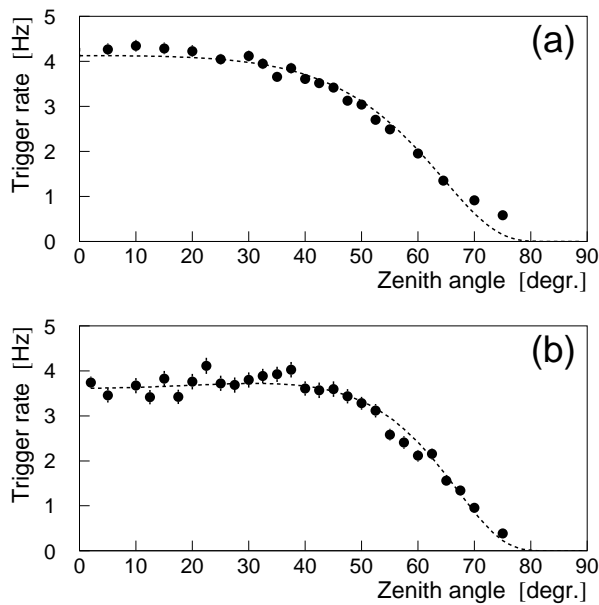}
}
\end{center}
\caption
{Single-telescope rate (a) and coincidence rate between 
telescopes CT3 and CT4 (b) as a function
of the zenith angle. The absolute values of the rates
cannot be compared, since data were taken with a higher
threshold in case of the single-telescope trigger.}
\label{fig_zenith}
\end{figure}

For future studies of systematics, in particular concerning
observations at large zenith angles, the dependence of 
single-telescope and multi-telescope cosmic-ray trigger
rates on the zenith angle
is of interest. Fig.~\ref{fig_zenith} shows the relevant data.
In case of the two-telescope coincidence, the telescopes
were always pointed perpendicular to the line connecting the 
telescopes.
Rates are relatively flat out to zenith angles of 
$40^\circ$. As the zenith angle increases, the distance
of showers to the telescope increases, reducing the density of
photons in the light pool and hence increasing the effective
energy threshold of the telescope. At the same time the diameter
of the light pool and the effective area increase.
The size of the images decreases, resulting in the concentration
of light in fewer pixels and in more efficient triggers.
For angles up to $40^\circ$, these effects more or less
compensate each other, resulting in a constant rate. At
larger angles, the increase in threshold - also due to
atmospheric extinction - dominates. A two-telescope coincidence
profits more from the enlarged light pool at non-zero zenith angles,
and the drop at larger zenith angles starts later,
compared to a single telescope.
Simple analytical models can only approximately describe this
behaviour; the dashed lines in Fig.~\ref{fig_zenith} represent
such an attempt.

\section{Trigger characteristics of $\gamma$--rays as compared to cosmic rays}

\begin{figure}[htb]
\begin{center}
\mbox{
\epsfxsize6.0cm
\epsffile{ 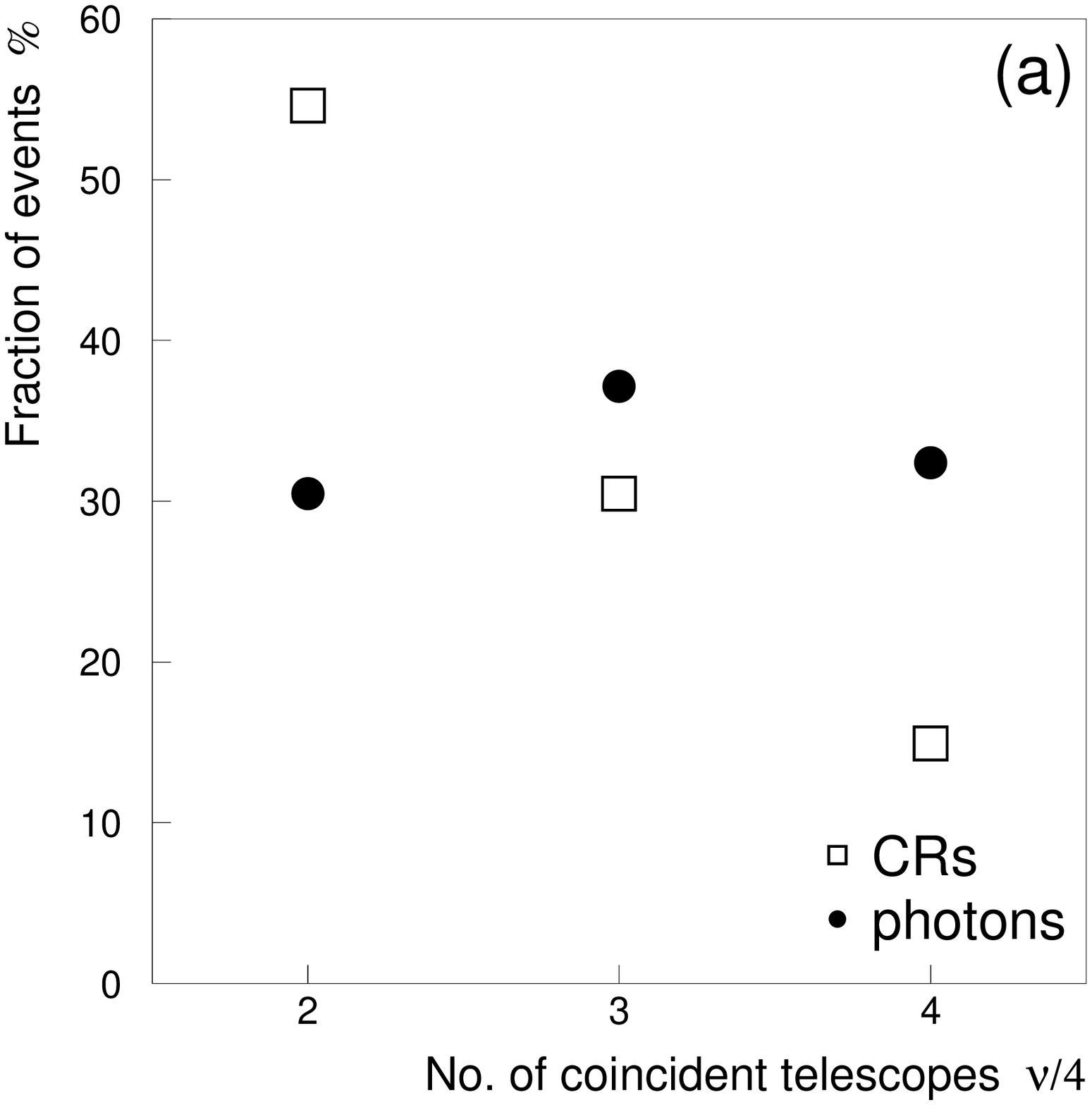}
}
\mbox{
\epsfxsize6.0cm
\epsffile{ 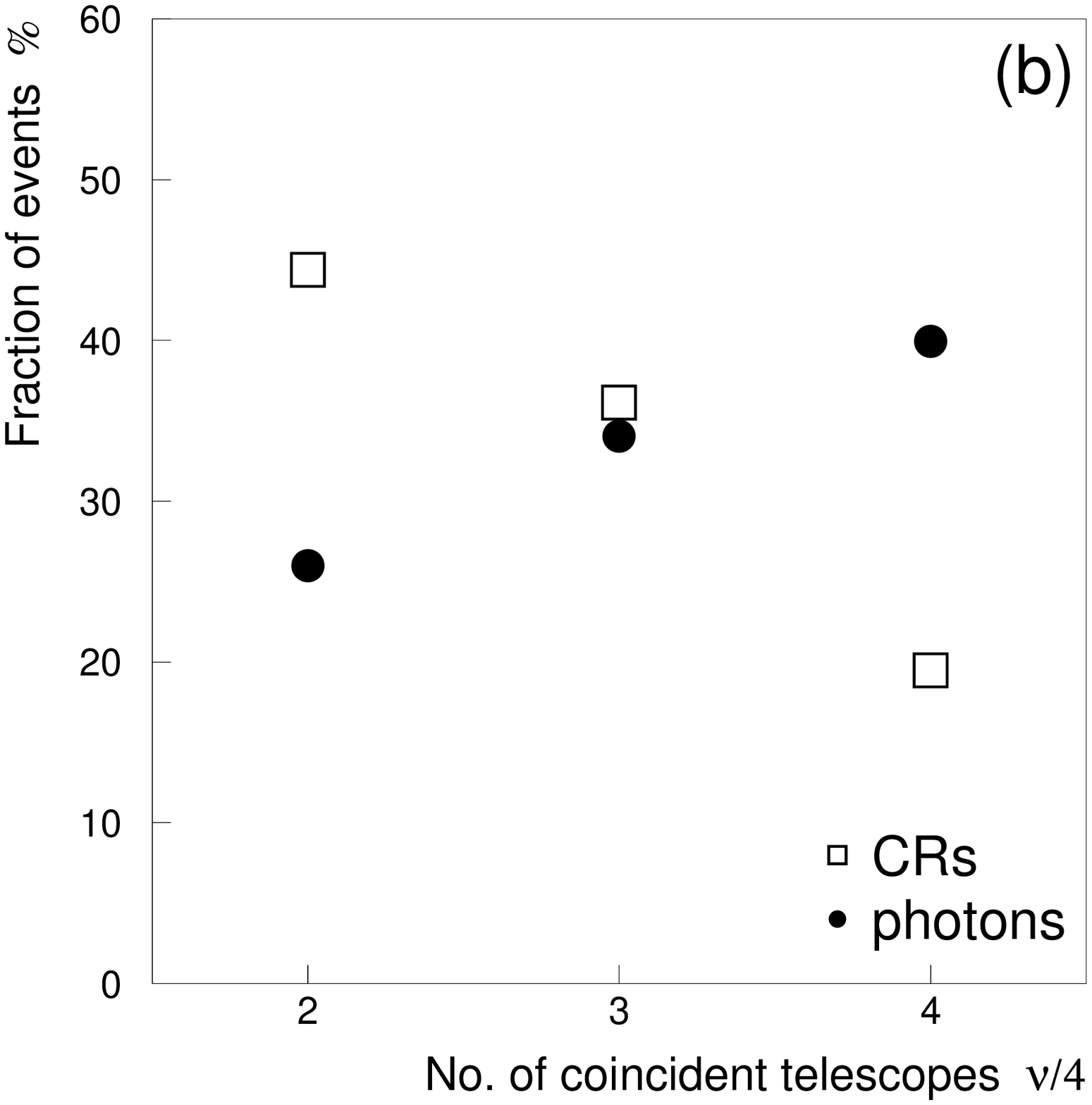}
}
\end{center}
\caption
{Fraction of two-, three-, and four-telescope events for
cosmic-ray triggers  (open squares) 
and for $\gamma$--rays from the Mkn 501 data
sample (full circles), for zenith angles between $12^\circ$ to 
$20^\circ$ (a) and between $30^\circ$ to 
$40^\circ$ (b).}
\label{fig_gamma}
\end{figure}
All measurements discussed above refer to cosmic-ray trigger
rates. Of course, it is interesting to see if
the trigger properties of $\gamma$--rays differ. Monte-Carlo
simulations, e.g., predict that a multi-telescope
coincidence suppresses cosmic rays already at the trigger
level~\cite{system_paper}. 
The recent flare of Mkn 501 has provided a large sample
a $\gamma$--rays, which can be clearly separated from cosmic
rays on the basis of their direction~\cite{system_501};
the small cosmic-ray background under the $\gamma$--ray signal
can be reliably subtracted on a statistical basis.

The flaring nature of the $\gamma$--ray emission of Mkn 501 does
unfortunately not allow to compare rates measured at 
different times, and hence does not allow to determine
unambiguously a zenith-angle dependence of $\gamma$-ray trigger rates.
We will therefore limit ourselves to the discussion of the
relative rates of twofold, threefold, and fourfold telescope coincidences.
In order to minimize a possible selection bias, $\gamma$--rays
were defined as showers within $0.32^\circ$ from Mkn 501
(compared to a typical angular resolution of $0.1^\circ$),
with a mean width of the images (averaged over all telescopes)
not more than 40\% larger than expected for $\gamma$--rays.
This latter cut is more than 95\% efficient. Mkn 501 data
were taken with the source located $0.5^\circ$ away from
the center of cameras; therefore, an opposite, but otherwise
equivalent region can be used to study background cosmic rays
under identical conditions. The resulting
relative coincidence rates are compared for $\gamma$--rays and
cosmic rays in Fig.~\ref{fig_gamma}, for two different 
zenith angles. One finds that $\gamma$-ray showers tend to
trigger a larger number of telescopes, as compared to 
cosmic-ray showers. The more telescopes are required at the
trigger level, the larger is the fraction of $\gamma$-rays
among the triggered events. With increasing zenith angles, the
fraction of four-telescope coincidences increases and the 
fraction of twofold coincidences decreases, as expected due
to the growing diameter of the light pool.

\section{Summary}

Using the HEGRA IACT system, single-telescope and multi-telescope
coincidence rates were studied as a function of pixel threshold
and telescope orientation, using both a standard two-pixel
coincidence to trigger an individual telescope, and a topological
trigger which requires neighboring trigger pixels.

The multi-telescope trigger scheme of the HEGRA IACT System allows
to operate the system at a pixel threshold lower by a factor of two
compared to single telescope operation with a standard two-pixel 
coincidence. At the system level the role of random coincidences
is negligible and therefore the limiting factor 
for further reduction of the energy threshold is the quality of the recorded
images. 

Studies with the Mkn 501 $\gamma$-ray sample demonstrate
that a multi-telescope trigger enhances the fraction of 
$\gamma$-rays in the event sample.

\section*{Acknowledgements}

The support of the HEGRA experiment by the German Ministry for Research 
and Technology BMBF and by the Spanish Research Council
CYCIT is acknowledged. We are grateful to the Instituto
de Astrofisica de Canarias for the use of the site and
for providing excellent working conditions. We thank the other
members of the HEGRA CT group, who participated in the construction,
installation, and operation of the telescopes. We gratefully
acknowledge the technical support staff of Heidelberg,
Kiel, Munich, and Yerevan.

\end{document}